\documentclass[pra,aps,superscriptaddress,twocolumn,nopacs,nofootinbib]{revtex4}
\usepackage{graphicx,color,epstopdf}
\usepackage{amsmath,amsfonts,enumerate,amsthm,amssymb,mathtools}
\usepackage{enumitem}
\usepackage{thmtools,thm-restate}
\usepackage{bbold}
\usepackage{hyperref}
\usepackage{multirow}
\usepackage{subfigure} % Include figure files
\usepackage{mathdots} %to use futher dots commands
\usepackage{float}

\newcommand\coolover[2]{\mathrlap{\smash{\overbrace{\phantom{%
    \begin{matrix} #2 \end{matrix}}}^{\mbox{$#1$}}}}#2}

\newcommand\coolunder[2]{\mathrlap{\smash{\underbrace{\phantom{%
    \begin{matrix} #2 \end{matrix}}}_{\mbox{$#1$}}}}#2}

\newcommand\coolleftbrace[2]{%
#1\left\{\vphantom{\begin{matrix} #2 \end{matrix}}\right.}

\newcommand\coolrightbrace[2]{%
\left.\vphantom{\begin{matrix} #1 \end{matrix}}\right\}#2}

\usepackage{blkarray}
\usepackage{kbordermatrix}

\hypersetup{
   colorlinks=true,         % false: boxed links; true: colored links, false is default
   linkcolor=blue,          % color of internal links, red is default
   citecolor=blue           % color of links to bibliography
}

\DeclareMathOperator{\Tr}{Tr}

\DeclarePairedDelimiter\floor{\lfloor}{\rfloor}

\theoremstyle{plain}

\theoremstyle{definition}

\theoremstyle{remark}

\definecolor{orange}{rgb}{1,0.5,0}
\definecolor{geen}{rgb}{0,0.8,0}

\begin{document}

\title{Quantum error correction codes and absolutely maximally entangled states}
\begin{abstract}
\end{abstract}

\author{Pawe\l{} Mazurek}
\affiliation{Institute of Theoretical Physics and Astrophysics, National Quantum Information Centre, Faculty of Mathematics, Physics and Informatics, University of Gda\'nsk, 80-308 Gda\'nsk, Poland}
\affiliation{International Centre for Theory of Quantum Information, University of Gda\'nsk, 80-308 Gda\'nsk, Poland}

\author{M\'at\'e Farkas}
\affiliation{Institute of Theoretical Physics and Astrophysics, National Quantum Information Centre, Faculty of Mathematics, Physics and Informatics, University of Gda\'nsk, 80-308 Gda\'nsk, Poland}

\author{Andrzej Grudka}
\affiliation{Faculty of Physics, Adam Mickiewicz University, 61-614 Pozna\'n, Poland}

\author{Micha\l{} Horodecki}
\affiliation{Institute of Theoretical Physics and Astrophysics, National Quantum Information Centre, Faculty of Mathematics, Physics and Informatics, University of Gda\'nsk, 80-308 Gda\'nsk, Poland}
\affiliation{International Centre for Theory of Quantum Information, University of Gda\'nsk, 80-308 Gda\'nsk, Poland}

\author{Micha\l{} Studzi\'nski}
\affiliation{Institute of Theoretical Physics and Astrophysics, National Quantum Information Centre, Faculty of Mathematics, Physics and Informatics, University of Gda\'nsk, 80-308 Gda\'nsk, Poland}

\begin{abstract}
For every stabiliser $N$-qudit absolutely maximally entangled state, we present a method for determining the stabiliser generators and logical operators of a corresponding quantum error correction code. These codes encode $k$ qudits into $N-k$ qudits, with $k\leq \floor{N/2}$, where the local dimension $d$ is prime. We use these methods to analyse the concatenation of such quantum codes and link this procedure to entanglement swapping. Using our techniques, we investigate the spread of quantum information on a tensor network code formerly used as a toy model for the AdS/CFT correspondence. In this network, we show how corrections arise to the Ryu-Takayanagi formula in the case of entangled input state, and that the bound on the entanglement entropy of the boundary state is saturated for absolutely maximally entangled input states.
\end{abstract}

\maketitle

\section{Introduction}

A powerful connection between theories of gravity and quantum field theory is the so-called AdS/CFT correspondence~\cite{Maldacena1999}. This relates gravitational theories in the interior (bulk) of anti-de Sitter (AdS) spaces to conformal field theories (CFT) defined on the boundary of the space. This duality allows for translating calculations from weakly interacting, and thus more tractable, gravitational models to strongly interacting quantum field theoretical models. A particularly interesting feature of this correspondence is the so-called Ryu-Takayanagi formula~\cite{Ryu2006}, which states that the entropy of a subregion of the boundary is proportional to the length of the geodesic connecting the endpoints of this subregion.

Recently the quantum information community is also focusing attention on the AdS/CFT correspondence. In~\cite{Pastawski2015}, the authors present a toy model of the correspondence, using a discrete tiling of the AdS space. They define a tensor on every tile, and every neighbouring tensor pair is connected by an edge, forming a network of tensors. On every edge, they define a quantum system of local dimension $d$. Interestingly, the Ryu-Takayanagi formula holds for every concise region of the boundary, as long as the above tensors are \emph{perfect tensors}. This means that if we treat the tensors as mappings between their edges, they define an isometry from any of their $n$ edges to any of their remaining $m$ edges, as long as $n\le m$.

Perfect tensors are important in quantum information theory, because they correspond to absolutely maximally entangled (AME) states, that is, states that are maximally entangled on every bipartition.
AME states are in close connection with quantum error correction codes~\cite{Scott2004}, and they are a necessary resource in multipartite quantum communication and quantum secret sharing schemes~\cite{Helwig2012}. From the mathematical perspective, many AME state constructions can be obtained from combinatorial designs~\cite{Goyeneche2015, Goyeneche2017}. 

In this work, we exploit the connection between AME states and quantum error correcting codes in order to gain understanding on the toy model of the AdS/CFT correspondence. We treat the perfect tensors of the network as AME states and therefore quantum codes, which allows us to describe the spread of the encoded quantum information in the network using code concatenation and the stabiliser formalism of quantum codes.

We derive the form of logical operators and stabiliser generators for every code emerging from an $n+m$ qudit stabiliser AME state, that encodes $n$ qudits into $m$ qudits with $n\leq m$. The proof is valid for local dimensions being prime, and is based on properties of graph states, into which every stabiliser AME state can be mapped under local unitaries. Apart from this, for a specific perfect tensor network, we analyse the spread of entanglement along the boundary of the code, with respect to the nature of input entanglement. We find that the upper bound of the entanglement entropy of the boundary state is saturated for AME input states for every connected bipartition of the boundary. We postulate that this optimality is another characteristic of AME states.

This paper is organised as follows: In Section \ref{Sec:2} we present the stabiliser formalism and describe the construction and applications of AME states, followed by the relation between stabiliser AME states and stabiliser quantum error correction codes. In Section \ref{Sec:3}, we describe how to construct tensor networks using quantum code concatenation and entanglement swapping. Then in Section \ref{Sec:4} we apply this procedure to analyse the spread of quantum information on a network used in the toy models of the AdS/CFT correspondence. Finally, in Section \ref{Sec:5} we apply our formalism to study the entanglement entropy of the boundary state of the perfect tensor network with respect to the initial input state.

\section{Stabliser codes and AME stabiliser states}\label{Sec:2}

We start by showing the connection between states that are maximally entangled on a bipartition, and quantum error correction codes. We assume that both the state and the code can be described within the stabiliser formalism, which we outline below.

\subsection{Stabiliser formalism}
Let us denote a Hilbert space of $N$ qudits by $\mathcal{H}=\otimes_{i=1}^{N} \mathcal{H}_{i}$, where $\mathcal{H}_{i}\cong\mathbb{C}^{d}$, and the space of linear operators acting on this Hilbert space by $\mathcal{B}(\mathcal{H})$. On this space, we  call a state $|\Phi\rangle$ a \emph{stabiliser state} if it is uniquely (up to a global phase factor) characterised by an Abelian subgroup $\mathcal{S}$ of the Pauli group $P_{N}\subset\mathcal{B}(\mathcal{H})$ via the equations
\begin{equation}\label{eq:stabiliser}
s|\Phi\rangle=|\Phi\rangle~~~\forall s\in \mathcal{S},
\end{equation}
that is, the group $S$ \emph{stabilises} $|\Phi\rangle$. The Pauli group $P_N$ is generated by $\{ \omega ^c X_j^{a_j} Z_j^{b_j} \}_{j = 1,\ldots,N}^{a_j, b_j, c = 0,\ldots,d-1}$, where $\omega = \mathrm{e}^{\frac{2 \pi \mathrm{i}}{d}}$ is the $d$-th root of unity, and $X_j$ and $Z_j$ are the generalised Pauli operators acting on the $j$-th qudit, defined as
\begin{equation}\label{eq:Pauli}
X = \sum_{k = 0}^{d - 1} | k + 1 \rangle \langle k | ~~~ Z = \sum_{k = 0}^{d - 1} \omega^k | k \rangle \langle k |,
\end{equation}
where, $| 0 \rangle, \ldots, | d - 1 \rangle$ is the computational basis.
Note that for any nontrivial stabiliser state we have that $\omega^c \mathcal{I}\notin\mathcal{S}$ if $c \neq 0$, where $\mathcal{I}$ is the identity operator on $\mathcal{H}$.

It is often convenient to describe the stabiliser group $\mathcal{S}$ by its generating set, $\mathcal{G(S)}$, which is a set of mutually commuting operators from $P_{N}$. If a group $\mathcal{S}\subseteq P_N$ does not define a unique state, but a higher dimensional subspace, we will call this subspace $\mathcal{H}_{log}=\{|\Phi\rangle: \forall s\in \mathcal{S}$, $s|\Phi\rangle=|\Phi\rangle\} $ a \emph{logical subspace}, and it defines a quantum code. In this case, there exist linearly independent operators from $P_{N}$ commuting with $\mathcal{S}$, but linearly independent from $\mathcal{S}$. We will call such operators \emph{logical operators} of the code, and, due to the fact that they commute with $\mathcal{S}$, we define them modulo $\mathcal{G(S)}$. Let $D=N-|\mathcal{G(S)}|$, where $|\cdot|$ denotes the cardinality of the set. Then we have that $\dim\mathcal{H}_{log}=d^{D}$, and $\mathcal{H}_{log}\cong\mathcal{H}_{L,1}\otimes\dots\otimes\mathcal{H}_{L,D}$, where $\mathcal{H}_{L,j}\cong\mathbb{C}^{d}$ is the $j$-th logical subsystem of the code. The set of operators commuting with $\mathcal{S}$ can then be written as $\{X_{L,1}, Z_{L,1},\dots,X_{L,D}, Z_{L,D}\}\cup\mathcal{S}$, where $(X_{L,j},Z_{L,j})$ form pairs of logical operators acting non-trivially only on $\mathcal{H}_{L,j}$, such that $Z_{L, j}X_{L, j} = \omega X_{L, j} Z_{L, j}$. Therefore, logical operators labeled with different indices commute with each other.
    
\subsection{AME states}

Consider a state $|\Phi\rangle$ on a Hilbert space of $N$ qudits: $\mathcal{H}=\otimes_{i=1}^{N} \mathcal{H}_{i}$, $\mathcal{H}_{i}\cong\mathbb{C}^{d}$. For the set of indices $I=\{1,\dots,N\}$ numbering the qudits, define a bipartition $I=A\cup B$ into two, non-empty sets of indices. We will call a state $|\Phi\rangle$  
\emph{maximally entangled} (ME) with respect to the bipartition $\mathcal{H}=\mathcal{H}_{A}\otimes\mathcal{H}_{B}$,
if it satisfies the following equivalent criteria (where, without loss of generality, we assume that $m\coloneqq|B|\leq \lfloor\frac{N}{2}\rfloor$):
\begin{enumerate}[label=(\roman*)]
\item $\Tr_{A}\big(|\Phi\rangle\langle\Phi|\big)=\mathcal{I}_{B}$
\item $S\big(\Tr_{A}|\Phi\rangle\langle\Phi|\big)=S(B)=|B|\log d$
\item $|\Phi\rangle=\frac{1}{\sqrt{d^{m}}}\sum_{k\in\mathbb{Z}^{m}_{d}}|k_{1}\rangle_{B_{1}}\dots|k_{m}\rangle_{B_{m}}|\Psi(k)\rangle_{A}$, 
with $\langle \Psi(k)|\Psi(k')\rangle=\delta_{k,k'}$.
\end{enumerate}
Above, $\Tr_{A}$ is the partial trace over $\mathcal{H}_{A}$, $S(\rho)=-\Tr\rho\log\rho$ is the von Neumann entropy, and $\delta_{m,n}$ is the Kronecker delta. If a state $|\Phi\rangle$ satisfies the above criteria for every bipartition, it is called absolutely maximally entangled (AME). From condition (i) we see that if a state is ME with respect to every bipartition such that $m=\lfloor\frac{N}{2}\rfloor$, then it is also AME.

For qubit systems ($d=2$), AME states exist for $N=2$ parties ($|\Psi\rangle\propto|00\rangle+|11\rangle$) and $N=3$ parties ($|\Psi\rangle\propto|000\rangle+|111\rangle$), but for $N=4$ qubits AME states do not exist~\cite{Higuchi2000}. By numerically minimizing the function $\sum_{B : |B|=\lfloor\frac{N}{2}\rfloor}\Tr \rho_{B}^{2}$, where $\rho_{B}=\Tr_{A}\rho$, AME states for $N=5,6$ qubits were found in~\cite{Facchi2008}. A recent result~\cite{Hubert2016}  showed that qubit AME states do not exist for $N= 7$, complementing the earlier result on their absence for $N\geq 8$~\cite{Scott2004}. On the other hand, for every $N$, there always exists large enough $d$ such that AME states exist~\cite{Helwig2012}.

An interesting application of AME states comes from their connection to isometries. From the defining equation (i) it can be easily seen that any decomposition of an AME state into two subsystems $|\Phi\rangle=\sum_{a,b}T_{a,b}|a\rangle|b\rangle$ is uniquely associated with a map $T: |b\rangle\rightarrow \sum_{a}T_{a,b}|a\rangle$ that satisfies $\sum_{a}T^{\dagger}_{b',a}T_{a,b}=\delta_{b',b}$, i.e.~that preserves all the inner products. That is, transformations associated with AME states are isometries with respect to every division into two subsystems, and are therefore perfect tensors.

This property is what allows AME-based tensor networks to be used as toy models for the AdS/CFT correspondence, because it ensures that an equivalent of the Ryu-Takayanagi formula holds in the network. However, we will show in Section~\ref{Sec:5} that this formula gains some corrections if the input state of the tensor network is entangled.

From now on, we will assume that the AME states corresponding to the tensors of the network are stabiliser states. To facilitate the description of networks in the stabiliser language, below we describe the stabilisers and logical operators of quantum codes corresponding to stabiliser AME states.

\subsection{Construction of stabiliser codes from AME states}

Every $N$-qudit stabiliser state can be mapped to a so-called graph state by Clifford operations acting locally on the qudits, whenever $d$ is prime~\cite{Schlingemann2001}. Since local Clifford operations do not affect entanglement between parties, every stabiliser AME state can be mapped to a graph state using local Cliffords. The generators of the stabiliser group of an $N$-qudit graph state are $N$ operators of the form $g_{i}= X_i \prod_{j = 1}^N Z_{j}^{A_{i,j}}$. Here, $X_i$ and $Z_j$ are generalised Pauli matrices acting on the $i$-th ($j$-th) qudit, and $A_{i,j}$ is the \emph{adjacency matrix} of a weighted graph, that is, a symmetric matrix with elements from the set $\{0,\ldots,d-1\}$, such that all its diagonal elements are 0.

Let $K=\{k_{1},\dots,k_{m}\}$ be any subset of the indices $\{1,\ldots,N\}$, indexing the generators (or the subsystems), such that $m\leq \floor*{\frac{N}{2}}$. We denote by $A_{i}/K$ the $i$-th row of the matrix $A$ with entries removed from the columns indexed by $K$. 
Theorem 7 of~\cite{Helwig2013} shows that, whenever $A$ is the adjacency matrix of an AME state, the set of vectors $\{A_{i}/K\}_{i\in K}$ is linearly independent in $\mathbb{Z}_d^{N-m}$. Let us choose $K$ to encompass the first $m$ rows/columns of $A$, that is, $k_{i}=i$.

Therefore, the first $m$ rows of $A$, truncated to their last $N-m$ entries, are linearly independent. Thus, if we construct a matrix of these truncated rows, we can perform Gauss-Jordan elimination to bring it to a reduced row echelon form. Then, by swapping the columns of $A$, we can move the $m$ columns containing just one `1' to the last $m$ positions. Since multiplication of the rows of the adjacency matrix $A$ corresponds to multiplication of the generators of the corresponding AME state, the above operations on $A$ can be directly translated into operations on the list of stabiliser generators: (1) Gauss-Jordan elimination, (2) reordering or columns: 

\begin{widetext}
\[
\begin{matrix}% matrix for left braces
    \coolleftbrace{N}{\\ \\ \\ \\ \\ \\ \\ \\ \\}\\
\end{matrix}%
\begin{bmatrix}
 \coolover{N}{X & \dots & \dots & \dots& \dots& \dots& \dots& \dots & \dots & \\}
\dots & X & \dots& \dots& \dots& \dots& \dots& \dots & \dots \\
\dots & \dots & X& \dots& \dots& \dots& \dots& \dots & \dots \\
\dots & \dots & \dots& X& \dots& \dots& \dots& \dots & \dots \\
\dots & \dots & \dots& \dots& X& \dots& \dots& \dots & \dots \\
\dots & \dots & \dots& \dots& \dots& X& \dots& \dots & \dots \\
\dots & \dots & \dots& \dots& \dots& \dots& X& \dots & \dots \\
\dots & \dots & \dots& \dots& \dots& \dots& \dots& X & \dots \\
\dots & \dots & \dots& \dots& \dots& \dots& \coolunder{m}{\dots& \dots & X &} \\
\end{bmatrix}%
\begin{matrix}% matrix for right braces
 \coolrightbrace{ \dots \\\dots \\X }{m}\\
\vphantom{a}\\
\vphantom{a}\\
\vphantom{a}\\
 \coolrightbrace{ \dots \\\dots \\X }{m}\\
   % \coolrightbrace{y \\ y \\ x }{U}
\end{matrix}
\xrightarrow{(1)}
\begin{bmatrix}
 \coolover{m}{X & \dots & X} & 1& Z& Z&\coolover{m}{ 1& 1 & \dots \phantom{a} & }\\
X & X & \dots& 1& 1& 1& Z& 1 & \dots \\
\dots & X & X& 1& 1& 1& 1& Z & Z \\
\dots & \dots & \dots& X& \dots& \dots& \dots& \dots & \dots \\
\dots & \dots & \dots& \dots& X& \dots& \dots& \dots & \dots \\
\dots & \dots & \dots& \dots& \dots& X& \dots& \dots & \dots \\
\dots & \dots & \dots& \dots& \dots& \dots& X& \dots & \dots \\
\dots & \dots & \dots& \dots& \dots& \dots& \dots& X & \dots \\
\dots & \dots & \dots& \dots& \dots& \dots& \coolunder{m}{\dots& \dots & X &} \\
\end{bmatrix}%
\begin{matrix}% matrix for right braces
 \coolrightbrace{ \dots \\\dots \\X }{m}\\
\vphantom{a}\\
\vphantom{a}\\
\vphantom{a}\\
 \coolrightbrace{ \dots \\\dots \\X }{m}\\
   % \coolrightbrace{y \\ y \\ x }{U}
\end{matrix}
\xrightarrow{(2)}
\]
\end{widetext}

\begin{widetext}
\[
\xrightarrow{(2)}
\begin{bmatrix}
 \coolover{m}{X & \dots & X} & 1& Z& \dots&\coolover{m}{ Z& 1 & 1\phantom{a}  & }\\
X & X & \dots& 1& 1& \dots& 1& Z & 1 \\
\dots & X & X& 1& 1& Z& 1& 1 & Z \\
\dots & \dots & \dots& X& \dots& \dots& \dots& \dots & \dots \\
\dots & \dots & \dots& \dots& \dots& \dots& X& \dots & \dots \\
\dots & \dots & \dots& \dots& X& \dots& \dots& \dots & \dots \\
\dots & \dots & \dots& \dots& \dots& \dots& \dots& X & \dots \\
\dots & \dots & \dots& \dots& \dots& \dots& \dots& \dots & X \\
\dots & \dots & \dots& \dots& \dots& X& \coolunder{m}{\dots& \dots & \dots &} \\
\end{bmatrix}%
\begin{matrix}% matrix for right braces
 \coolrightbrace{ \dots \\\dots \\X }{m}\\
\vphantom{a}\\
\vphantom{a}\\
\vphantom{a}\\
 \coolrightbrace{ \dots \\\dots \\X }{m}\\
   % \coolrightbrace{y \\ y \\ x }{U}
\end{matrix}
\xrightarrow{(3)}
\begin{bmatrix}
 \coolover{m}{X & \dots & X} & \dots& \dots& \dots&\coolover{m}{ Z& 1 & 1 \phantom{a} & }\\
X & X & \dots& \dots& \dots& \dots& 1& Z & 1 \\
\dots & X & X& \dots& \dots& \dots& 1& 1 & Z \\
\dots & \dots & \dots& X& \dots& \dots& \dots& \dots & \dots \\
\dots & \dots & \dots& \dots& X& \dots& \dots& \dots & \dots \\
\dots & \dots & \dots& \dots& \dots& X& \dots& \dots & \dots \\
\dots & \dots & \dots& \dots& \dots& \dots& X& \dots & \dots \\
\dots & \dots & \dots& \dots& \dots& \dots& \dots& X & \dots \\
\dots & \dots & \dots& \dots& \dots& \dots& \coolunder{m}{\dots& \dots & X &} \\
\end{bmatrix}%
\begin{matrix}% matrix for right braces
 \coolrightbrace{ \dots \\\dots \\X }{m}\\
\vphantom{a}\\
\vphantom{a}\\
\vphantom{a}\\
 \coolrightbrace{ \dots \\\dots \\X }{m}\\
   % \coolrightbrace{y \\ y \\ x }{U}
\end{matrix}
\xrightarrow{(4)}\\
\]
\end{widetext}
Above, every row corresponds to a different generator, every column to a qudit, and an entry with possible $1$ or $Z^a$ assignment is marked by ``$\dots$''.  
Note that swapping qudits in (2) may destroy the diagonal structure of the $X$ operators. However, by proper swapping of rows from $m+1$ to $N$, we can recover this structure in step (3).

Finally, by properly multiplying the last $N-m$ generators by the first $m$ generators (4), we can completely eradicate the $Z$ operators from the last $m$ positions of the last $N-m$ generators:

\begin{widetext}
\[
\xrightarrow{(4)}
\begin{bmatrix}
 \coolover{m}{X & \dots & X} & \dots& \dots& \dots&\coolover{m}{ Z& 1 & 1 \phantom{a} & }\\
X & X & \dots& \dots& \dots& \dots& 1& Z & 1 \\
\dots & X & X& \dots& \dots& \dots& 1& 1 & Z \\
\dots & \dots & \dots& \dots& \dots& \dots& 1& 1 & 1 \\
\dots & \dots & \dots& \dots& \dots& \dots&1& 1 & 1 \\
\dots & \dots & \dots& \dots& \dots& \dots& 1& 1 & 1 \\
\dots & \dots & \dots& \dots& \dots& \dots& X& 1 & 1 \\
\dots & \dots & \dots& \dots& \dots& \dots& 1& X & 1 \\
\dots & \dots & \dots& \dots& \dots& \dots& \coolunder{m}{1& 1 & X &} \\
\end{bmatrix}%
\begin{matrix}% matrix for right braces
 \coolrightbrace{ \dots \\\dots \\X }{m}\\
\vphantom{a}\\
\vphantom{a}\\
\vphantom{a}\\
 \coolrightbrace{ \dots \\\dots \\X }{m}\\
   % \coolrightbrace{y \\ y \\ x }{U}
\end{matrix}
\]
\end{widetext}

Note that here, in some places marked by ``$\dots$'' in the first $m$ columns of the last $N-m$ rows, $X$ or $XZ$ operators may also be present. Generally speaking, we obtain a list of $N$ generators with a fixed structure on the last $m$ qudits (full separation and local diagonalisation with respect to $Z$ and $X$ operators), and possibly mixed entries on the remaining qudits.

We will show that these $N$ generators truncated to the first $N-m$ qudits provide the generators and logical operators of an error correction code encoding $m$ qudits into $N-m$ qudits. For this, we need to show that all the operators are linearly independent, and we have $N-2m$ commuting operators providing the stabiliser generators. Moreover, the remaining $2m$ operators form pairs satisfying the same commutation relations as $X$ and $Z$ Pauli operators, and they commute with the $N-2m$ generators.

We prove linear independence of the truncated operators by contradiction: assume that the truncated list is not linearly independent. Then, by multiplying some generators with each other, we are able to obtain a truncated generator that consists only of identity operators. However, this contradicts with~\cite[Theorem 7]{Helwig2013}, which states that any $m$ generators of an AME state truncated to length $N-m$ are linearly independent (and this property is preserved under multiplying the generators with each other).

To see commutativity, notice that the truncated generators from rows $m+1,\dots,N-m-1$ of the above matrix commute with each other, as the full generators (that commute with each other) act trivially on the last $m$ positions that were truncated. These truncated generators are therefore stabiliser generators of an error correction code. On the other hand, the truncated operators from the first and last $m$ rows commute with the newly established generators of the code, and for every truncated operator from the first $m$ rows, there is exactly one truncated operator from the last $m$ rows with which it commutes the same way as Pauli X and Z operators.

Also notice that each of these operators commutes with all other truncated operators apart from their pairs. Therefore, the truncated operators from the first and last $m$ rows form logical operators of the above quantum error correction code. As $m\in\{1,\dots,\floor*{\frac{N}{2}}\}$, every stabilizer AME state of $N$ qubits generates at least $\floor*{\frac{N}{2}}$ different stabiliser codes. Each of these codes encodes $m$ logical qudits into $N-m$ physical ones.

We will illustrate the connection between AME states and quantum error correction codes by an example of a 6-qubit AME state~\cite{Borras2007}, characterised by the following list of stabiliser generators: 
\begin{widetext}
\[
\begin{bmatrix}
X & Z & Z & X& 1& 1  \\ 
1 & X & Z & Z& X& 1  \\
X & 1 & X & Z& Z& 1  \\
Z & X & 1 & X& Z& 1  \\
X & X & X & X& X& X  \\
Z & Z & Z & Z& Z& Z  \\\end{bmatrix}
\begin{matrix}
 1\\
 2\\
 3\\
 4\\
 5\\
 6\\
\end{matrix}
\longrightarrow
\begin{bmatrix}
- Y & Z & Y & 1& 1& Z  \\
-Z & X & Z & 1& 1& X  \\
Y & Y & Z & 1& Z& 1  \\
Z & Z & X & 1& X& 1  \\
-Z & Y & Y & Z& 1& 1  \\
X & Z & Z & X&1& 1  \\\end{bmatrix}
\begin{matrix}
 3\cdot 6\\
 2\cdot 3\cdot 4\cdot 5\\
 1\cdot 4\\
 1\cdot 2\cdot 3\cdot 4\\
 1\cdot 3\cdot 4\\
 1\\
\end{matrix}\]
\begin{equation}\label{AME}
{\color{white}}
\end{equation}
\end{widetext}

Here, even without referring to local Clifford transformations, we can obtain a desired, alternative list of stabiliser generators by multiplying them with each other (the numbers on the right side of each new generator denote which original generators have to be multiplied to obtain it). From the reasoning presented above for the general case, it follows that the obtained form explicitly expresses stabiliser generators and logical operators for $1\rightarrow 5$, $2\rightarrow 4$, and $3\rightarrow 3$ codes. Namely, for the $1\rightarrow 5$ case we have
\begin{equation}
\mathcal{G(S)}=\{YYZ1Z,ZZX1X,-ZYYZ1,XZZX1\}
\end{equation}
and
\begin{equation}
Z_{L,1}=-YZY11, ~~~ X_{L,1}=-ZXZ11,
\end{equation}
for the $2\rightarrow 4$ case we get
\begin{equation}
\mathcal{G(S)}=\{-ZYYZ,XZZX\}
\end{equation}
and
\begin{equation}
\begin{split}
& \left. Z_{L,1}=-YZY1,~~~X_{L,1}=-ZXZ1, \right. \\
& \left. Z_{L,2}=YYZ1,~~~X_{L,2}=ZZX1, \right.
\end{split}
\end{equation} 
and finaly for the $3\rightarrow 3$ case we obtain
\begin{equation}
\mathcal{G(S)}=\emptyset
\end{equation}
and
\begin{equation}
\begin{split}
& \left. Z_{L,1}=-YZY,~~~X_{L,1}=-ZXZ, \right. \\
& \left. Z_{L,2}=YYZ,~~~X_{L,2}=ZZX, \right. \\
& \left. Z_{L,3}=-ZYY,~~~X_{L,3}=XZZ. \right.
\end{split}
\end{equation}

\section{Concatenation of codes and entanglement swapping between states}\label{Sec:3}

Having achieved the above standard form of the list of stabiliser generators, we are ready to describe a particular measurement process performed on AME states. It will enable us to understand the construction of perfect tensor networks, and associated error correction codes, as a concatenation of codes corresponding to AME states.

First, let us recall the rules of updating the list of generators of a stabiliser state after performing a measurement~\cite[Chapter 10.5]{NielsenChuang}. If the measured observable can be constructed from the generators, then we do not need to update the list. If the measured observable cannot be constructed as a product of generators, but commutes with all of them, then we add it to the list of generators with a phase factor depending on the measurement outcome. Finally, if the observable cannot be constructed as a product of generators and does not commute with some number of generators, then it replaces one generator with which it does not commute (possibly multiplied by a phase factor depending on the measurement outcome), while we multiply the other non-commuting generators by the generator that we removed from the list -- this process assures that all the new generators are linearly independent and mutually commuting. 

Using these rules and our previous machinery, we provide a connection between concatenation of stabiliser codes and entanglement swapping. Let us start with two stabiliser states, which, by the position of their stabiliser generators on the common stabiliser list (see below), we will call \emph{left} and \emph{right} states, and denote by $L$ and $R$, respectively. The left state is defined on $N_{L}$ qudits and is arbitrary, while the right one is defined on $N_{R}$ qudits, and we take it to be an AME state. Each stabiliser of the $L$ state can be represented in the form $L_{i}\otimes \sigma_{i}$, where $L_{i}$ is a tensor product of Pauli operators on qudits labeled from $l_{1}$ to $l_{N_{L}-1}$, while $\sigma_{i}$ is a single qudit Pauli operator on qudit $l_{N_L}$, and $i=1,\dots,N_{L}$. Similarly, every stabiliser of the state $R$ is of the form $\sigma_{j}\otimes R_{j}$, with $R_{j}$ acting on qudits $r_{2},\ldots,r_{N_{R}}$, $\sigma_j$ acting on qudit $r_1$, and $j=1,\dots,N_{R}$. The list of $N_{L}+N_{R}$ stabilisers of the joint state, being a product of the left and right states, takes the form presented in the first table below:

\begin{widetext}  
\[
\begin{blockarray}{ccccccccccc}
\begin{block}{(ccccccccccc)}
  & & L_{1}  &  X & 1 & &\dots&  1  \\
  & & {\color{red}L_{2} } & {\color{red} Z} & {\color{red}1} & &{\color{red}\dots}& {\color{red} 1}  \\
  & & L_{3}  &  1 & 1 & &\dots&  1  \\
  & & L_{4}  &  X & 1 & &\dots&  1  \\
 & &  & &  \dots & & &    \\
   & & {\color{red}L_{N_{L}} }& {\color{red}Z} &{\color{red} 1} & &{\color{red}\dots}& {\color{red} 1}  \\
  {\color{red}1} & & {\color{red}\dots}   & {\color{red}1} & {\color{red}Z} & &   {\color{red} R_{1}}&    \\
  1 & & \dots   & 1 & X & &    R_{2}&    \\
  1 & & \dots   & 1 & 1 & &    R_{3}&    \\
  1 & & \dots   & 1 & 1 & &    R_{4}&    \\
 & &  & &  \dots & & &    \\
  1 & & \dots  & 1 & 1 & &    R_{N_{R}}&    \\
\end{block}
   l_{1} & l_{2}&  \dots&l_{N_{L}}   &   r_{1} & r_{2}&  \dots&r_{N_{R}}    \\
\end{blockarray}
\xrightarrow{XX}
\begin{blockarray}{ccccccccccc}
\begin{block}{(ccccccccccc)}
  & & {\color{red}L_{1} } &  {\color{red}X} & {\color{red}1} & &\dots& {\color{red} 1}  \\
  & & {\color{blue} L_{2} } & {\color{blue} Z} & {\mathbf{\color{blue}Z}} & &{\mathbf{\color{blue}R_{1}}}&   \\
  & & L_{3}  &  1 & 1 & &\dots&  1  \\
  & &{\color{red}L_{4}}  & {\color{red} X} &{\color{red}1} & &\dots&  {\color{red}1}  \\
 & &  & &  \dots & & &    \\
   & & {\color{blue}L_{N_{L}} }& {\color{blue} Z}   &  {\mathbf{\color{blue} Z}}  & &{\mathbf{\color{blue} R_{1}}}&    \\
  {\color{blue}1} & & {\color{blue}\dots }  &{\color{blue} X} &{\color{blue} X} & &    \dots& {\color{blue}1}   \\
  {\color{red}1} & & \dots   &{\color{red} 1 }& {\color{red}X} & &   {\color{red} R_{2}}&    \\
  1 & & \dots   & 1 & 1 & &    R_{3}&    \\
  1 & & \dots   & 1 & 1 & &    R_{4}&    \\
 & &  & &  \dots & & &    \\
  1 & & \dots  & 1 & 1 & &    R_{N_{R}}&    \\
\end{block}
  l_{1} & l_{2}&  \dots&l_{N_{L}}   &   r_{1} & r_{2}&  \dots&r_{N_{R}}    \\
\end{blockarray}
\xrightarrow{ZZ}
\begin{blockarray}{ccccccccccc}
\begin{block}{(ccccccccccc)}
  & & {\color{blue}L_{1} } &  {\color{blue}X} & {\mathbf{\color{blue}X}} & &\mathbf{\color{blue} R_{2}}&    \\
  & & {\color{black} L_{2} } & {\color{black} Z} & {\color{black}Z} & &{\color{black}R_{1}}&   \\
  & & L_{3}  &  1 & 1 & &\dots&  1  \\
  & & {\color{blue}L_{4} } &  {\color{blue}X} & \mathbf{\color{blue}X} & &\mathbf{\color{blue} R_{2}}&    \\
 & &  & &  \dots & & &    \\
 & & {\color{black}L_{N_{L}} }& {\color{black} Z}   &  {\color{black} Z}  & &{\color{black} R_{1}}&    \\
    {\color{black}1} & & {\color{black}\dots }  &{\color{black} X} &{\color{black} X} & & {\color{black}   \dots}& {\color{black}1}   \\
  {\color{blue}1} & & {\color{blue}\dots }  &{\color{blue} Z} &{\color{blue} Z} & & {\color{blue}   \dots}& {\color{blue}1}   \\
  1 & & \dots   & 1 & 1 & &    R_{3}&    \\
  1 & & \dots   & 1 & 1 & &    R_{4}&    \\
 & &  & &  \dots & & &    \\
  1 & & \dots  & 1 & 1 & &    R_{N_{R}}&    \\
\end{block}
   l_{1} & l_{2}&  \dots&l_{N_{L}}   &   r_{1} & r_{2}&  \dots&r_{N_{R}}    \\
\end{blockarray}
\xrightarrow{}\]
\[\xrightarrow{}
\begin{blockarray}{ccccccccccc}
\begin{block}{(ccccccccccc)}
& & &  \smash{\vrule width 0.5pt depth 140pt height 1pt }& \smash{\vrule width 0.5pt depth 140pt height 1pt }\\
  & & {\color{black}L_{1} } &  {\color{black}1} & {\color{black}1} & &{\color{black} R_{2}}&    \\
  & & {\color{black} L_{2} } & {\color{black} 1} & {\color{black}1} & &{\color{black}R_{1}}&   \\
  & & L_{3}  &  1 & 1 & &\dots&  1  \\
  & & {\color{black}L_{4} } &  {\color{black}1} & {\color{black}1} & &{\color{black} R_{2}}&    \\
 & &  & &  \dots & & &    \\
 & & {\color{black}L_{N_{L}} }& {\color{black} 1}   &  {\color{black} 1}  & &{\color{black} R_{1}}&    \\
 \hbox to 0pt{\vrule width 1.8in depth -3pt height 3.5pt\hss}   {\color{black}1} & & {\color{black}\dots }  &{\color{black} X} &{\color{black} X} & & {\color{black}   \dots}& {\color{black}1}   \\
 \hbox to 0pt{\vrule width 1.8in depth -3pt height 3.5pt\hss}  {\color{black}1} & & {\color{black}\dots }  &{\color{black} Z} &{\color{black} Z} & & {\color{black}   \dots}& {\color{black}1}   \\
  1 & & \dots   & 1 & 1 & &    R_{3}&    \\
  1 & & \dots   & 1 & 1 & &    R_{4}&    \\
 & &  & &  \dots & & &    \\
  1 & & \dots  & 1 & 1 & &    R_{N_{R}}&    \\
\end{block}
  l_{1} & l_{2}&  \dots&l_{N_{L}}   &   r_{1} & r_{2}&  \dots&r_{N_{R}}    \\
\end{blockarray}
\rightarrow
\begin{blockarray}{ccccccccccc}
\begin{block}{(ccccccccccc)}
  & & {\color{magenta}L_{1} }     & & &{\color{magenta} R_{2}}&    \\
  & & {\color{magenta}L_{2} }     & & &{\color{magenta} R_{1}}&    \\
  & & {\color{magenta}L_{3}}  &  {\color{magenta}1} & {\color{magenta}1}  &{\color{magenta}\dots}& {\color{magenta} 1}  \\
  & & {\color{magenta}{\color{magenta}L_{4} }} &   &   &{\color{magenta}{\color{magenta} R_{2}}}&    \\
 & &  & &  {\color{black}\dots} & & &    \\
 & & {\color{magenta}L_{N_{L}} }&      & &{\color{magenta} R_{1}}&    \\
  {\color{orange}1} & & {\color{orange}\dots}   & {\color{orange}1} &  &     {\color{orange}R_{3}}&    \\
  {\color{orange}1} & & {\color{orange}\dots}   & {\color{orange}1 }&  &    {\color{orange}R_{4}}&    \\
 & &  & &  \dots & & &    \\
  {\color{orange}1 }& & {\color{orange}\dots}  & {\color{orange}1 }&  &    {\color{orange}R_{N_{R}}}&    \\
\end{block}
  l_{1} & l_{2}&  \dots&l_{N_{L-1}}    & r_{2}&  \dots&r_{N_{R}}    \\
\end{blockarray}
\]
\end{widetext}

Note that in the first table we already exploited the fact that the $R$ state is AME. Therefore, its stabiliser generators can be written such that on an arbitrarily selected qudit there are only two generators acting non-trivially: one as $X$, and the other as $Z$. In the first step, we measure the observable $XX$ on qudits $l_{N_{L}}$ and $r_{1}$. A non-commuting generator in the seventh line of the table is replaced by the measured operator (we assume here that the measurement outcome was $+1$), and the other non-commuting generators from the second and sixth lines are multiplied by the removed generator from the seventh line (all marked with red in the first table). 

In the next step, we measure the observable $ZZ$ on the same pair of qudits and apply the same update procedure (the newly modified generators are marked with blue, the generator in the eighth line will be replaced by the operator being measured, while the first and fourth generators will be multiplied by it, all marked with red). At the end, we see that the qudits labeled by $l_{N_{L}}$ and $r_{1}$ are maximally entangled with each other, and are correlated with no other part of the system. We trace them out by erasing the two corresponding columns and rows of the table.

The resulting table, representing the stabiliser generators of the state formed in the process of entanglement swapping on a pair of qudits shared between the state $L$ and the AME state $R$, consists of two groups: the original stabilisers of the state $R$, trivially extended two the whole system (marked in orange), and the stabilisers of the state $L$, multiplied by truncated stabilisers of the state $R$ (marked in purple). Note that this multiplication is conditioned on the $l_{N_{L}}$-th component of the stabiliser of the $L$ state: if it is $X$ ($Z$), the stabiliser will be multiplied by $R_{1}$ ($R_{2}$). Due to the initial AME structure of the state $R$, this construction can be extended to the case of arbitrary number of $(XX,ZZ)$ measurements performed on up to $\min \{\floor{N_{L}/2},\floor{N_{R}/2}\} $, arbitrarily selected, disjoint pairs of qudits.

Noting that $R_{1}$ and $R_{2}$ are the logical $X$ and $Z$ operators of the code described by the AME state $R$, encoding 1 qudit into $N_{R}-1$ qudits, we see that entanglement swapping of AME states is equivalent to concatenating codes described by these states. In the most basic case, by entanglement swapping performed between a single qudit state and an arbitrary qudit of an AME state, we encode the single qudit state into the logical subspace of the stabiliser code, that lives on the remaining qudits of the AME state.

\begin{figure}
\includegraphics[scale=0.4]{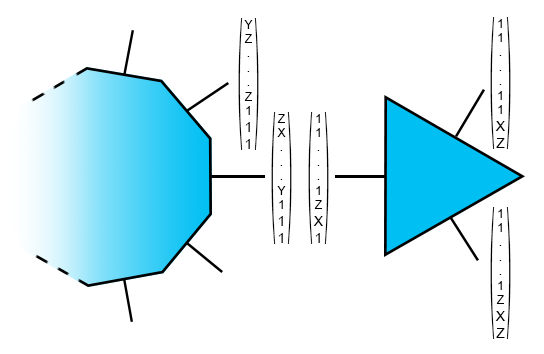}
\caption{\label{tri1} A multiqubit state (left) vs. an AME state, specifically a 3-qubit GHZ state (right). Vectors corresponding to every uncontracted leg (a qubit) are taken from the corresponding column of the matrix formed by the list of generators of the two states. Therefore, different entries of a given vector describe the action of different stabilisers on the corresponding qubit.}
\end{figure}

\begin{figure}
\includegraphics[scale=0.4]{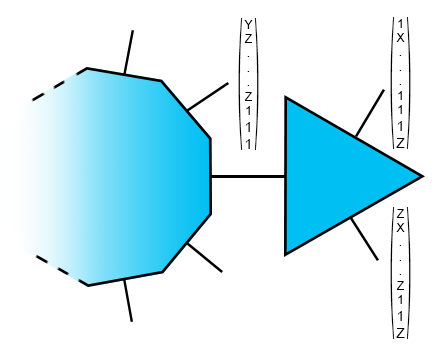}
\caption{\label{tri2} After measuring the adjacent qubits in the Bell basis (contracting the leg connecting the polygon with the triangle), the state on the uncontracted legs (qubits) is fully described by the updated generators. Note that the generators of the polygon do not change (apart from the one associated with the contracted leg), while the generators formerly belonging to the AME state get updated according to the vectors of the qubits that are contracted. }
\end{figure}

\begin{figure}
\includegraphics[scale=0.35]{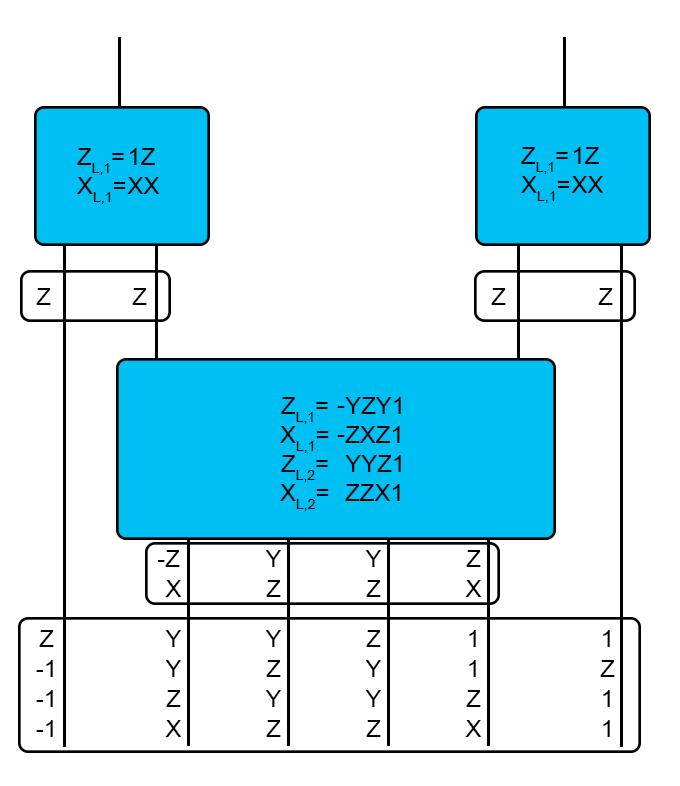}
\caption{\label{concat1} Example of the emergence of stabilizers through concatenation. Generators are given in transparent frames and logical operators are given in blue frames. From the top: 2 codes, based on 3-qubit GHZ states, each encoding one qubit into two qubits. The stabiliser generator of the resulting codes is $ZZ$. Then, 2 of the resulting qubits get encoded into a 4 qubit code, based on 6-qubit AME state. The generators of this code appear. On the other hand, the previously existing stabilisers get concatenated through logical operators of the 4-qubit code, and the final list of stabilizers is given in the bottom frame. Thus this total set of stabilisers consists of two classes: (i) those in the first two lines,
which are "old" stabilizers, just encoded through the 4 qubit code (ii) those in the last two lines -- the "new" ones -- which are the stabilisers of the 
4-qubit code. }	
\end{figure}

The above code concatenation procedure is visualised on Figs. \ref{tri1}, \ref{tri2} and \ref{concat1}. On Figs. \ref{tri1} and \ref{tri2}, we use codes emerging from the 3-qubit GHZ state, and the 6-qubit AME state in Eq. \eqref{AME}. In the first figure, we have a pair of quantum states: a polygon on the left, representing a state from previous stages of concatenation, and therefore not necessarily AME, and a triangle representing a 3-qubit GHZ state on the right. After entanglement swapping on a pair of qubits, the vectors describing how stabilisers act on individual qubits get updated (Fig. \ref{tri2}). Fig. \ref{concat1} shows how the stabilisers of the final state emerge during the concatenation of three codes.

Some stabilisers result from encoding previous stabilisers through logical operators, while others appear locally when they are not concatenated further.

If the state $L$ is AME as well, then the list of stabiliser generators of the state resulting from entanglement swapping
is composed simply from the original generators of states $R$ and $L$, acting non-trivially only within their original domains, and two generators being the product of logical $X$ and $Z$ operators: $Z_{L,1}\otimes Z_{L,2}$ and $X_{L,1}\otimes X_{L,2}$ (truncated to qubits which do not belong to the qubit pair on which the measurements were performed). Clearly, such a state does not have to be AME.   

\section{Spread of quantum information in AME-based networks}\label{Sec:4}

\begin{figure}
\includegraphics[scale=0.3]{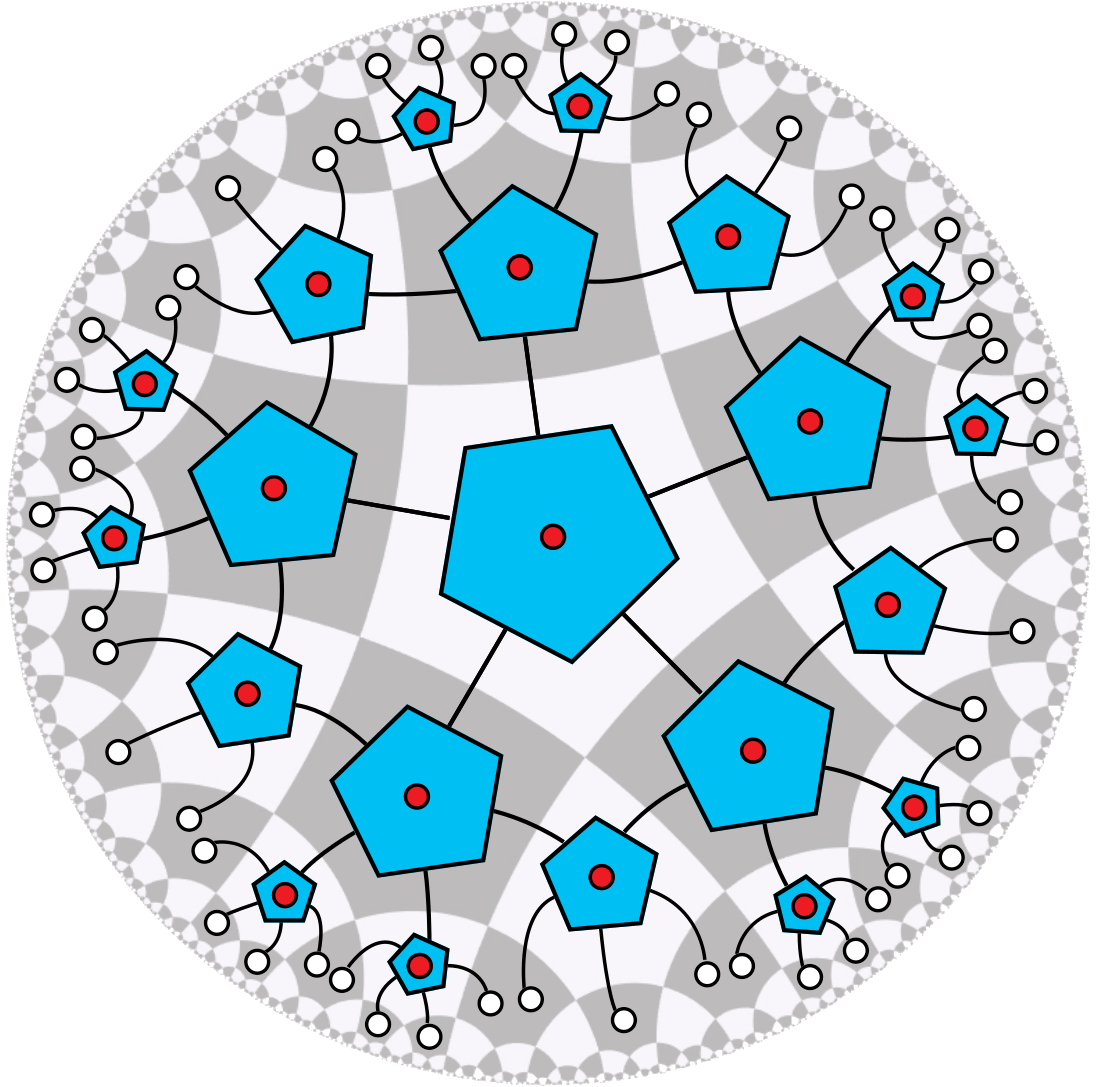}
\caption{\label{Fig:network} Tiling of a surface with negative curvature into pentagons.  Red dots inside every pentagon, as well as the white dots coming out of the pentagons, represent qubits. Each of the pentagons represents the 6-qubit AME state. Contracted legs represent entanglement swapping. }
\end{figure}

Having connected entanglement swapping with the concatenation procedure, we are now able to exploit the stabiliser formalism to investigate the spread of quantum information in perfect tensor networks based on stabiliser AME states. The final code can be formed by a gradual expansion of the network through entanglement swapping, where in each step at least one of the states is AME. Note that we can extend the procedure presented in the previous section to describe the case when more than two AME states are the subject of entanglement swapping. Due to the restriction $\min \{ \floor{N_{L}/2},\floor{N_{R}/2} \} $ of the number of qubits of a given AME state that can participate in this gradual concatenation, the networks that accommodate the spread of quantum information must to have a non-decreasing number of branches, starting from the point of the origin. Here, we analyse a network investigated in~\cite{Pastawski2015}, based on a tiling of a two dimensional space with negative curvature (see Fig. \ref{Fig:network}). The tiling consists of pentagons, which are connected by edges, representing qubits. Each of the pentagons is associated with the 6-qubit AME state, described in Eq. \eqref{AME}. The association is that each of the legs coming from an edge of a pentagon corresponds to a qubit state, while the sixth qubit is associated with a leg that emerges from the center of the pentagon, and stretches above the plane in which the figure lies.

\begin{figure}
\includegraphics[scale=0.45]{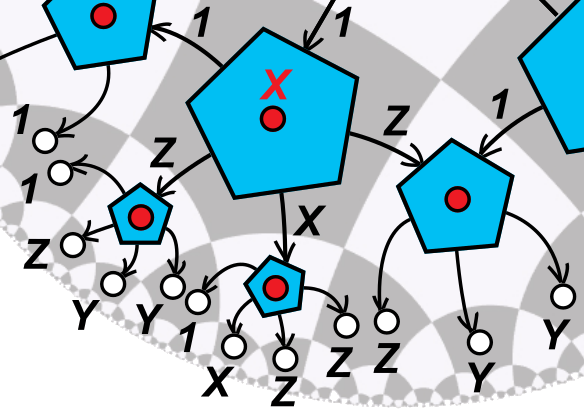}
\caption{\label{Fig:zoom} Spread of the operator $X$ applied to the red entry (logical qudit) of the largest pentagon. First, it gets encoded in the form of the logical operator $X_1 = -ZXZ1$ of the $1 \to 5$ code emerging from a 6-qubit AME state. Further concatenation then results in the final form of the operator on the boundary state.}
\end{figure}

In most cases, we will treat the interior red dots as logical qudits, and the outer white dots as physical qudits, encoding the logical qudits into the code space of the emerging quantum code.
The encoding of the operator $X$ from one of the interior points of the networks (logical qudit) to the boundary (physical qudits) can be understood in the following way (see Fig.~\ref{Fig:zoom}): an $X$ operator that acts originally on a logical qudit, gets encoded as a result of entanglement swapping between qudit pairs of consecutive AME states of the tensor network, and the encoded operators on each level are determined by the logical operators of the corresponding quantum codes.

Whether $1\rightarrow 5$, $2\rightarrow 4$ or $3\rightarrow 3$ logical operators are used for a selected pentagon depends only on the number of inputs and outputs of the pentagon. In order to determine which legs are inputs and outputs of certain tensors, we assign a direction to each leg, pointing ``away'' from the origin of the space, and we treat every red dot as an input.

At the end of the concatenation process, the form of the encoded $X$ operator is given modulo stabilisers, that are defined by the concatenation as well. Note that the final form of the logical operators and generators results from concatenation, that can be viewed in terms of performing commuting measurements (entanglement swapping on disjoint qudit pairs). Therefore, the order of the concatenation, governed by the direction of arrows, can only influence the final form of logical operators and generators up to multiplication by stabilisers, i.e.~it cannot have any measurable consequences.   

\section{Entanglement corrections to the Ryu-Takayanagi formula}\label{Sec:5}

\begin{figure}
\includegraphics[scale=0.9]{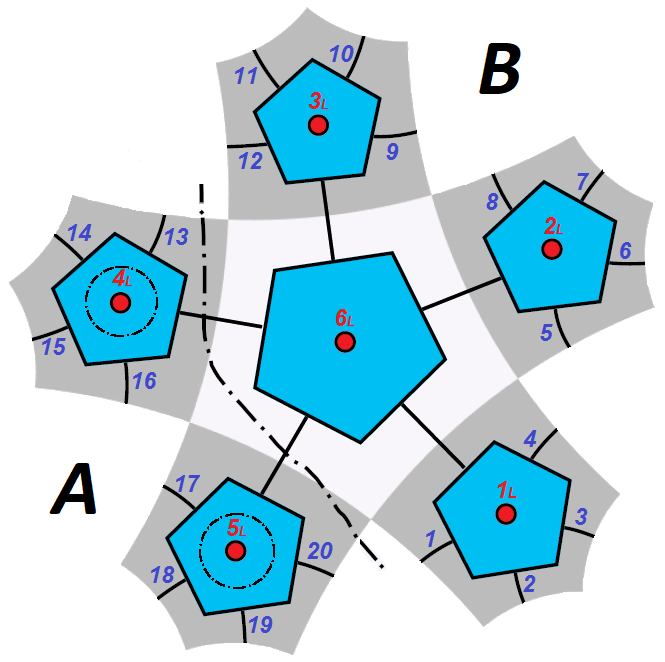}
\caption{\label{cor} Labeling of logical qubits (red dots inside pentagons) and physical qubits (non-contracted links on the boundary) for studies of entropy of entanglement of the encoded states.  The region $A$ consists of qubits labeled by $1,\ldots, s_B-1$, while the region $B$ consists of qubits labeled by $13,\ldots,s_B$. Here, $s_{B}=13$.}
\end{figure}
In this section we will apply our techniques to analyse corrections 
to the Ryu-Takayanagi formula in the pentagon code 
in~\cite{Pastawski2015}. Namely, as noted in~\cite{Pastawski2015}, if the input state is 
product, then the Ryu-Takayanagi formula holds, but when it is entangled it may not hold anymore. Here we provide a simple bound on the entanglement entropy of 
a connected region of the boundary, and check whether it is saturated by various input states for the network in Fig.~\ref{cor}, which has 2 layers of concatenation. 

We find that among the selected several states, only AME states saturate the bound (\ref{bound}). 

Let us recall the Ryu-Takayanagi formula, which states that if we divide the boundary of the network into two disjoint but connected regions $A$ and $B$, then the entanglement entropy of the boundary state along this cut is
\begin{equation}\label{eq:Ryu}
S_A \propto | \gamma_A |,
\end{equation}
where $\gamma_A$ is a geodesic in the interior of the network, connecting the two points where $A$ and $B$ meet. The ``volume'' of this geodesic, $|\gamma_A|$, in our case is simply the number of legs the geodesic cuts through.

In general, the state on the boundary can be described as
\begin{align}\label{eq:bipartite}
|\Psi\rangle=\sum\limits_{i,a,b,k,k'}\psi_{k,k'} P_{a,i}^{(k)}|a\rangle_{A}Q_{k,i}^{(k')}|b\rangle_{B}=\nonumber\\
\sum\limits_{k,k'}\psi_{k,k'}\sum\limits_{i}|P_{i}^{(k)}\rangle_{A}|Q_{i}^{(k')}\rangle_{B},
\end{align}
where $A$ and $B$ are the above mentioned disjoint connected regions with some bases $\{|a\rangle\}$ and $\{|b\rangle\}$ defined on them. Any path $\gamma$ that connects the two points on the boundary where $A$ and $B$ meet divides the operator corresponding to the full tensor network into two operators, $P$ and $Q$. The index $i$ runs through the qudits this path cuts through, whereas $k$ and $k'$ run through the (red) input qudits of $P$ and $Q$. Therefore, these operators are defined via
$P: |i \rangle| k\rangle \rightarrow \sum\limits_{a} P_{a,i}^{(k)}|a \rangle$ and $Q: |i \rangle| k'\rangle \rightarrow \sum\limits_{a} Q_{b,i}^{(k')}|b\rangle$. Note that in Eq. \eqref{eq:bipartite}, we define $|P_{i}^{(k)}\rangle_{A} = \sum_a P_{a,i}^{(k)}|a\rangle_{A}$ and $|Q_{i}^{(k')}\rangle_{B} = \sum_b Q_{k,i}^{(k')}|b\rangle_{B}$.

When the input state $\sum_{k,k'}\psi_{k,k'}|k\rangle|k'\rangle=|k_{*}\rangle|k_{*}^{'}\rangle$ is product with respect to the bipartition into inputs of $P$ and $Q$, by adjusting the basis we can simply write
\begin{align}
|\Psi\rangle=\sum\limits_{i}|\tilde{P}_{i}\rangle_{A}|\tilde{Q}_{i}\rangle_{B},
\end{align}
where $|\tilde{P}_{i}\rangle=|P_{i}^{k_{*}}\rangle$ and  $|\tilde{Q}_{i}\rangle=|Q_{i}^{k_{*}^{'}}\rangle$. Now, if $P$ and $Q$ are isometries, we have $\langle P_{i}^{k}| P_{i'}^{k'}\rangle=\delta_{i,i'}\delta_{k,k'}=\langle Q_{i}^{k}| Q_{i'}^{k'}\rangle$, and therefore $\langle \tilde{P}_{i}| \tilde{P}_{i'}\rangle=\delta_{i,i'}=\langle \tilde{Q}_{i}| \tilde{Q}_{i'}\rangle$. Therefore, $\Tr_{B} |\Psi\rangle\langle \Psi| \propto \mathcal{I}_{A}$, and the entanglement entropy is the logarithm of the rank of the state (which is $d^{|\gamma|}$), i.e. 
\begin{equation}\label{eq:productbound}
S(\Tr_{B}|\Psi\rangle\langle \Psi|) = \log_2(d) \cdot |\gamma|,
\end{equation}
which is precisely the Ryu-Takayanagi formula if we minimise over $\gamma$. If $P$ and $Q$ are not isometries, the above constitutes an upper bound on the entropy. In the general case, when the input state might be entangled, we have 
\begin{align}
\Tr_{B} |\Psi\rangle\langle \Psi| =\sum\limits_{k,k',i,\tilde{k},\tilde{k'},\tilde{i}}\psi_{k,k'}\psi_{\tilde{k},\tilde{k'}}\langle Q_{i}^{k'}|Q_{\tilde{i}}^{\tilde{k'}}\rangle | P_{i}^{k} \rangle\langle P_{\tilde{i}}^{\tilde{k}}|,
\end{align}
and therefore the entropy of the reduced state can surpass the product state bound, as we get 
\begin{equation}\label{bound}
S(\Tr_{B}|\Psi\rangle\langle \Psi|)\leq \log_2(d) \cdot \left( |\gamma|+|P| \right),
\end{equation} 
where $|P|$ is the number of input states of $P$. The above bound is valid for any choice of the path $\gamma$, and therefore we choose it to minimise $|\gamma|+|P|$ (or just $|\gamma|$ for the product case, in which case $\gamma = \gamma_A$ is a geodesic connecting the two points where $A$ and $B$ meet).

For the encoding setting presented in Fig. \ref{cor}, we investigate the entropy of the reduced state of the boundary with respect to different bipartitions (qubit 20 vs.~the rest, qubits 19 and 20 vs.~the rest, and so on), and different input states (6-qubit AME state, 6-qubit GHZ state, one singlet pair (otherwise product), and a fully product state). We use the methods from the previous section to generate the list of stabilisers of the code on the boundary, as well as the encoded version of generators that stabilise the code state: 
\begin{equation}\label{AMElist}
\tiny
\begin{bmatrix}
X & Y & X & 1 & X & Y & X & 1 & Y & Z & Y & 1 & 1 & 1 & 1 & 1 & Y & Z & Y & 1\\
Y & Z & Y & 1 & Y & Z & Y & 1 & Z & X & Z & 1 & 1 & 1 & 1 & 1 & Z & X & Z & 1\\
Y & Z & Y & 1 & X & Y & X & 1 & X & Y & X & 1 & Y & Z & 1 & 1 & 1 & 1 & 1 & 1 \\
Z & X & Z & 1 & Y & Z & Y & 1 & Y & Z & Y & 1 & Z & X & Z & 1 & 1 & 1 & 1 & 1 \\
Z & Y & Y & Z & 1 & 1 & 1 & 1 & 1 & 1 & 1 & 1 & 1 & 1 & 1 & 1 & 1 & 1 & 1 & 1 \\
X & Z & Z & X & 1 & 1 & 1 & 1 & 1 & 1 & 1 & 1 & 1 & 1 & 1 & 1 & 1 & 1 & 1 & 1 \\
1 & 1 & 1 & 1 & Z & Y & Y & Z & 1 & 1 & 1 & 1 & 1 & 1 & 1 & 1 & 1 & 1 & 1 & 1 \\
1 & 1 & 1 & 1 & X & Z & Z & X & 1 & 1 & 1 & 1 & 1 & 1 & 1 & 1 & 1 & 1 & 1 & 1 \\
1 & 1 & 1 & 1 & 1 & 1 & 1 & 1 & Z & Y & Y & Z & 1 & 1 & 1 & 1 & 1 & 1 & 1 & 1 \\
1 & 1 & 1 & 1 & 1 & 1 & 1 & 1 & X & Z & Z & X & 1 & 1 & 1 & 1 & 1 & 1 & 1 & 1 \\
1 & 1 & 1 & 1 & 1 & 1 & 1 & 1 & 1 & 1 & 1 & 1 & Z & Y & Y & Z & 1 & 1 & 1 & 1 \\
1 & 1 & 1 & 1 & 1 & 1 & 1 & 1 & 1 & 1 & 1 & 1 & X & Z & Z & X & 1 & 1 & 1 & 1 \\
1 & 1 & 1 & 1 & 1 & 1 & 1 & 1 & 1 & 1 & 1 & 1 &  1 & 1 & 1 & 1 &Z & Y & Y & Z  \\
1 & 1 & 1 & 1 & 1 & 1 & 1 & 1 & 1 & 1 & 1 & 1 & 1 & 1 & 1 & 1 & X & Z & Z & X \\
%\\
1 & 1 & Z & X & 1 & 1 & Z & X & X & X & X & X & X & 1 & Y & Y & X & 1 & Y & Y \\
1 & X & Z & Z & X & 1 & X & Z & X & 1 & X & Z & 1 & X & Z & Z & 1 & 1 & 1 & 1 \\
X & 1 & Y & Y & X & X & X & X & 1 & 1 & Z & X & 1 & 1 & Z & X & X & 1 & Y & Y \\
Y & Y & X & Z & Y & Z & Y & 1 & Y & Y & X & Z & Z & Z & Z & Z & Y & Z & Y & 1 \\
X & X & X & X & X & X & X & X & X & X & X & X & X & X & X & X & X & X & X & X \\
Z & Z & Z & Z & Z & Z & Z & Z & Z & Z & Z & Z & Z & Z & Z & Z & Z & Z & Z & Z \\
\end{bmatrix}
\end{equation}

Above, columns correspond to qubits on the boundary, with the leftmost column corresponding to qubit 1, and the rightmost column to qubit 20. The first 14 rows are the stabiliser generators of the boundary state, 
while the last 6 rows are the encoded versions of the generators of the 6-qubit AME state (LHS of (\ref{AME})).

For different input states, the last 6 rows are replaced with encoded versions of the respective stabiliser generators: for the 6-qubit GHZ state these will correspond to the encoded versions of $XXXXXX$, $ZZ1111$, $1ZZ111$, $11ZZ11$ , $111ZZ1$, and $1111ZZ$. For the ``singlet pair'' input state, we take the encoded versions of $111XX1$, $111ZZ1$, $X1111$, $1X1111$, $11X111$ and $11111X$ (qubits $4_L$ and $5_L$ are entangled), while for a product state we take the encoded versions of $X1111$, $1X1111$, $11X111$, $111X11$, $1111X1$ and $11111X$. Above, we take the order of the input qubits to be $1_{L},\dots,6_{L}$ from Fig. \ref{cor}. 

The entanglement of the reduced state can be easily deduced from the list of the generators~\cite{Fattal2004}, via 
\begin{equation}
S(\Tr_{B}|\Phi\rangle\langle\Phi|)=\frac{|S_{AB}|}{2},
\end{equation}  
where $|S_{AB}|$ is the minimal number of generators acting non-trivially on both $A$ and $B$.

This minimisation is over all possible representations of the generators of the stabiliser group. From~\cite{Fattal2004}, we see that this representation can be found in the following way: Let us denote by $P_B$ the projection mapping $g_{A}\otimes g_{B}$ to $\mathcal{I}_{A}\otimes g_{B}$, and the list of generators by $\{g_i\}_i$. If the set $\{P_B (g_i) \}_i$ contains $n$ pairs of operators that do not commute with each other, but otherwise all operators commute, then we have
$|S_{AB}| = 2 n$. 

The above representation can be achieved by manipulating the initial generator list, such as (\ref{AMElist}): For a selected cut $A:B$, we check the commutation of every element with all other elements on the subsystem $B$. If two elements do not commute, then they will form a non-commuting pair. Then, we update the list by multiplying all the elements that do not commute with one element of the pair by the other element of the pair. For example, the list of the generators of the state stabilized by (\ref{AMElist}), and for region $B$ comprising qubits 18, 19 and 20, can be brought to the following form on the last three qubits:
\begin{equation}\label{AMElist2}
\tiny
\begin{bmatrix}
1 & Y & Y\\
X & Z & 1\\
%\\
Z & Y & 1\\
Z & X & X\\
%\\
Y & X & Z\\
Z & Y & Y\\
%\\
1 & 1 & 1\\ 
1 & 1 & 1\\
1 & 1 & 1\\
1 & 1 & 1\\
1 & 1 & 1\\
1 & 1 & 1\\
1 & 1 & 1\\
1 & 1 & 1\\
1 & 1 & 1\\
1 & 1 & 1\\
1 & 1 & 1\\
1 & 1 & 1\\
1 & 1 & 1\\
1 & 1 & 1\\
\end{bmatrix},
\end{equation}
from which we infer
\begin{equation}
S(\Tr_{B=\{18,19,20\}}|AME\rangle_{enc}\langle AME|_{enc})=3.
\end{equation}

\begin{figure}
\includegraphics[scale=0.22]{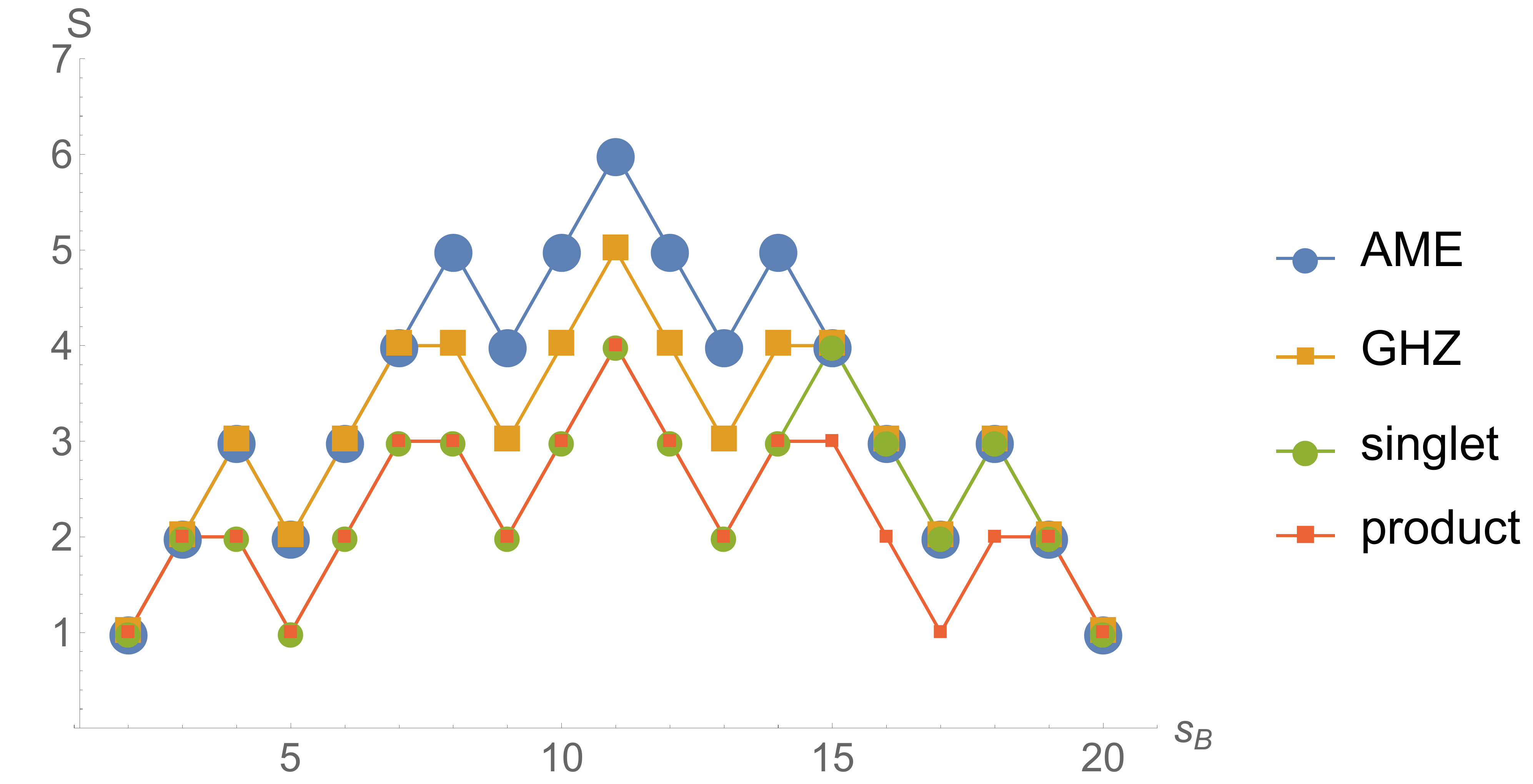}
\caption{\label{Entropy} 
Entanglement entropy $S(\Tr_{B}|\Psi\rangle\langle\Psi|)$ of the boundary states of the code from Fig. \ref{cor} for different input states (AME, GHZ, singlet or product, see the text for details), and for boundary points of the region $B$ being qubits $s_B$ and 20.
}
\end{figure}

We present the value of the entanglement entropy for different cuts and different encoded states in Fig. \ref{Entropy}. 

We see that for an arbitrary cut the AME state and the product state give limiting values of the entanglement entropy. The increased values of entropy (compared to product state input) for the ``singlet state'' input are present only for cuts that leave the entangled qubits, $4_{L}$ and $5_{L}$, on the opposite sides of the shortest path. However, we observe the difference between entanglement entropy even in cases when the same amount of entangled qubits is left on the opposite side of the optimal cut, i.e.~between encoded GHZ and AME states.
 For a particular cut, the bound \eqref{bound} is set by the geometry of the problem (Fig.~\ref{cor}), and one can check that the boundary state encoding the AME state saturates it. For example, for an AME input state and the division of the boundary determined by  $s_B=13$, the shortest path in the bulk is shown in Fig. \ref{cor} by a dash-dotted line. Dash-dotted circles surround logical inputs that account for the correction to the standard Ryu-Takayanagi formula.
 
Note that the corrections to the Ryu-Takayanagi formula correspond to the entanglement entropy of the input state with respect to the bipartition defined by the geodesic. This is maximised by AME states, therefore we conjecture that AME input states saturate the bound.

\section{Conclusions}
We have shown that each stabilizer AME state on $N$ qudits leads to a quantum error correction code encoding $k$ logical qudits into $N-k$ physical qudits, whenever $k=0,1,\dots,\floor{N/2}$. We provide an algorithm to determine the stabiliser generators and logical operators of the emerging code. We have also shown that entanglement swapping between two states such that one of them is AME enables to easily calculate the spread of quantum information in AME-based tensor networks, by concatenating stabilisers and logical operators.

Moreover, we have shown that the Ryu-Takayanagi formula acquires corrections in the case of tensor network codes with entangled inputs. In the particular case of the pentagon code with two layers of concatenation, we report that the bound on the corrections to the formula is saturated by AME input states for any bipartition of the boundary. We conjecture that AME input states saturate the entropy bound for boundary states of perfect tensor structures.

\textbf{Acknowledgments.} 
A.G., M.H.~and P.M.~are supported by the John Templeton Foundation through the grant ID \#56033. The opinions expressed in this publication are those of the authors and do not necessarily reflect the views of the John Templeton Foundation.
M.F.~is supported by the Polish NCN grant Sonata UMO-2014/14/E/ST2/00020. M.S.~is supported by the grant ``Mobilno\'s\'c Plus IV'', 1271/MOB/IV/2015/0 from the Polish
Ministry of Science and Higher Education.

\bibliography{Extremal_biblio}

\end{document}